\documentclass[amsmath,amssymb,aps,floatfix,prb,superscriptaddress,twocolumn]{revtex4-1}
\UseRawInputEncoding

\usepackage{graphicx}% Include figure files
\usepackage{xcolor}
\usepackage[version=4]{mhchem}
\usepackage{ulem}
\usepackage{hyperref}

\usepackage{graphicx}
\begin{document}

\title{Crystal Nucleation in Eutectic Al-Si Alloys by Machine-Learned Molecular Dynamics
%: a molecular dynamics study with machine-learning interatomic potential
}

\author{Quentin Bizot}

\email{Quentin.Bizot@grenoble-inp.fr}
\affiliation{Univ. Grenoble Alpes, CNRS, Grenoble INP, SIMaP, F-38000 Grenoble, France}

\author{Noel Jakse}
\affiliation{Univ. Grenoble Alpes, CNRS, Grenoble INP, SIMaP, F-38000 Grenoble, France}
\date{October 2024}

\begin{abstract}

%Aluminium alloyed with silicon is particularly interesting due to its properties, such as good corrosion resistance, excellent weldability, and low density. 
%These advantages, combined with the abundance and low cost of silicon, make this alloy widely used in various applications. Furthermore, the Al-Si binary is characterized by a eutectic in the Al rich part, with no stable intermetallic compounds across the entire composition range.
Solidification control is crucial in manufacturing technologies, as it determines the microstructure and, consequently, the performance of the final product. 
Investigating the mechanisms occurring during the early stages of nucleation remains experimentally challenging as it initiates on nanometer length and sub-picoseconds time  scales. 
Large scale molecular dynamics simulations using machine learning interatomic potential with quantum accuracy appears the dedicated approach to complex, atomic level, multidimensional mechanisms with local symmetry breaking. 
A potential trained on a high-dimensional neural network on density functional theory-based \textit{ab initio} molecular dynamics (AIMD) trajectories for liquid and undercooled states for Al-Si binary alloys enables us to study the nucleation mechanisms occurring at the early stages from the liquid phase near the eutectic composition.
Our results indicate that nucleation starts with Al in hypoeutectic conditions and with Si in hypereutectic conditions. 
Whereas Al nuclei grow in a globular shape, Si ones grow  with  polygonal faceting, whose underlying mechanisms are further discussed. 
\end{abstract}	
\maketitle

\section{Introduction}

Al-Si alloys offer several advantages such as low density, high wear resistance, low thermal coefficient and good mechanical strength at high temperatures. 
Its good casting ability makes it a material of choice for a wide range of industrial applications namely automotive, aeronautics, electronics, metallurgy \cite{singhal_review_2024,dash_review_2023}. 
Al-Si based alloys are also processed using a wide variety of manufacturing techniques, both traditional and additive manufacturing \cite{dash_review_2023}.
Despite these remarkable characteristics, these alloys present significant challenges when it comes to understanding their microstructural evolution, particularly during the solidification process. 
They do not exhibit the formation of stable crystalline compounds throughout the entire composition range. 
Instead, it features a pronounced eutectic point in the aluminum-rich region at 12.17 at\% Si \cite{livreAlSi_2016}. 
The system undergoes a phase separation into two phases: $\alpha$-Al (aluminium solid) with fcc structure and Si (silicon solid) with diamond structure \cite{nikanorov_structural_2005}. 
Depending on the composition, these phases may form lamellar or globular structures, which influence the alloy’s final mechanical and thermal properties \cite{zhang_research_2023}.

Understanding crystal nucleation and subsequent solidification mechanisms require precise insight into local atomic structure, dynamics as well as thermodynamic properties, which dictate the microstructure formation. 
From an experimental point of view, X-ray scattering and neutron diffraction measurements of structural and dynamical properties are very scarce, especially for the liquid phase, which is due to poor contrast between aluminum and silicon atoms \cite{mathiesen_x-ray_2011}. 
This is particularly true for diffusion, an important quantity for solidification control, for which no measurements exists.  
%In addition, thermodynamic quantities for the liquid phase are much debated in the literature \cite{mateiko_thermodynamic_2011} with a strong dependence to the temperature due to the presence of Si clusters or by a changing of the nature of bonds.
Moreover, early stages of crystal nucleation remains extremely challenging, and basically still out-of-reach of experimental investigation for this system.

In this context, atomic scale simulations emerge as a powerful alternative. Many studies using \textit{ab initio} Molecular Dynamics (AIMD) based on the density functional theory (DFT) aim at calculating various quantities of interest, such as structural, thermodynamic, and dynamic properties with a high degree of reliability \cite{qin_structure_2016, manga_ab_2018, fahs_structure-dynamics_2023}. 
Although predicting atomic interaction very accurately, DFT-based simulations are computationally intensive, and therefore limited to a few hundred atoms, leading typically to cell sizes of the order of the nanometer scale. This obviously restricts their ability to capture the larger-scale phenomena, notably those involved in solidification.
Classical Molecular Dynamics (MD) simulations enable to track nucleation events at atomic scales, providing critical data on cluster formation, and early-stage growth mechanisms under varying thermodynamic conditions for monatomic systems and alloys \cite{shibuta_homogeneous_2015, mahata_understanding_2018, orihara2020molecular, becker_unsupervised_2022, bizot_molecular_2023, shang2023influence, bizot_directional_2024, sandberg_homogeneous_2024}, including Al-Si alloys \cite{mahata_bridging_2024, liu_molecular_2016,li_molecular_2012, huang_liquid_2019, huang_improved_2018}. Nevertheless, the use of these classical potentials often limits the accuracy of the various properties as compared to experiments.

The emergence of the concept of Machine Learning Interatomic Potentials (MLIPs) \cite{behler2021four, unke2021machine, jacobs2025practical} makes it possible to combine the accuracy of \textit{ab initio} calculations and large-scale MD. 
They capture interactions by learning, through a supervised regression task, potential energy surface, forces, and stresses from local atomic environments.
Since the pioneering work introduced by Behler and Parinello on artificial neural networks (ANN) \cite{behler_generalized_2007}, several machine learning architectures have been developed, such as linear potential \cite{thompson2015spectral, cusentino2020explicit}, Gaussian Approximation Potentials (GAP) \cite{bartok_gaussian_2010}, Moment Tensor Potentials \cite{shapeev2016moment, hodapp2020operando}, High-Diensional Neural Network Potentials (HDNNP) \cite{behler2021four}, Graph Neural Networks \cite{batzner_e3-equivariant_2022}, and Deep Neural Network Potential \cite{zeng_deepmd-kit_2023} to mention a few.
Among these schemes, ANN potentials are inherently non-linear due to the presence of hidden layers and the use of non-linear activation functions between them. 
In most cases, non-linearity allows to take into account complex situations such as liquids with significant fluctuation of local environment along the phase-space trajectory during the simulation. 
%It offers a major advantage for neural network potentials to adequately study the liquid phase or the phase transition occurring during solidification.
A few number of studies start to appear that successfully demonstrated the ability of MLIPs to handle crystal nucleation of pure element such as Si and Al \cite{bonati_silicon_2018, miao_liquid_2020, jakse_machine_2022}, molecular systems \cite{piaggi2022homogeneous}, and in aluminum alloys \cite{sandberg_homogeneous_2024, zhai_accurate_2023}. For Al-Si alloys, a MLIP scheme was developed very recently within the GAP scheme but dedicated only to a very limited range of compositions \cite{liu_construction_2025}. 
%It is therefore essential to develop a general purpose MLIP transferable to all compositions for this alloy.

In this work, a general purpose MLIP is developed within the HDNNP framework with the aim of studying crystal nucleation in Al-Si alloys under hypo-eutectic and hyper-eutectic conditions. 
The potential is trained on AIMD trajectory data from the Al-Si alloy, covering the entire range of compositions, including pure Si and Al. 
The dataset includes liquid states at temperatures above and below the liquidus line, along with key crystalline phases at low temperatures for the pure elements.
The potential is successfully validated by assessing the microscopic structure with respect to X-ray diffraction experiments \cite{kazimirov_x-ray_2013}, and diffusion properties by a comparison with AIMD simulations. 
Large-scale MD simulations are then carried out eutectic nucleation under different conditions: homogeneous nucleation for a hypo-eutectic composition with 5 percent Si as well as heterogeneous conditions by seeding the liquid alloys pure Al and Si particles in their most stable structures for hyper- and hypo-eutectic compositions. 
Detailed observations are made on nucleation and phase growth behavior according to composition or particle type and a discussion of these findings with respect to experimental observations is given.

The layout of the paper is the following. Section \ref{sec:methods} describes the construction of the MLIP, the AIMD simulations to build the dataset as well as the classical MD simulations and relevant physical properties. Section \ref{sec:results} is devoted to the validation of the MLIP as well as the analysis of the crystal nucleation. The conclusions are provided in Section \ref{sec:conclusion}.

\section{Simulation background}
\label{sec:methods}

\subsection{Dataset: Ab-Initio Molecular Dynamics trajectories}

\begin{figure} 
    \centering 
    \includegraphics[angle=0, width=0.5\textwidth]{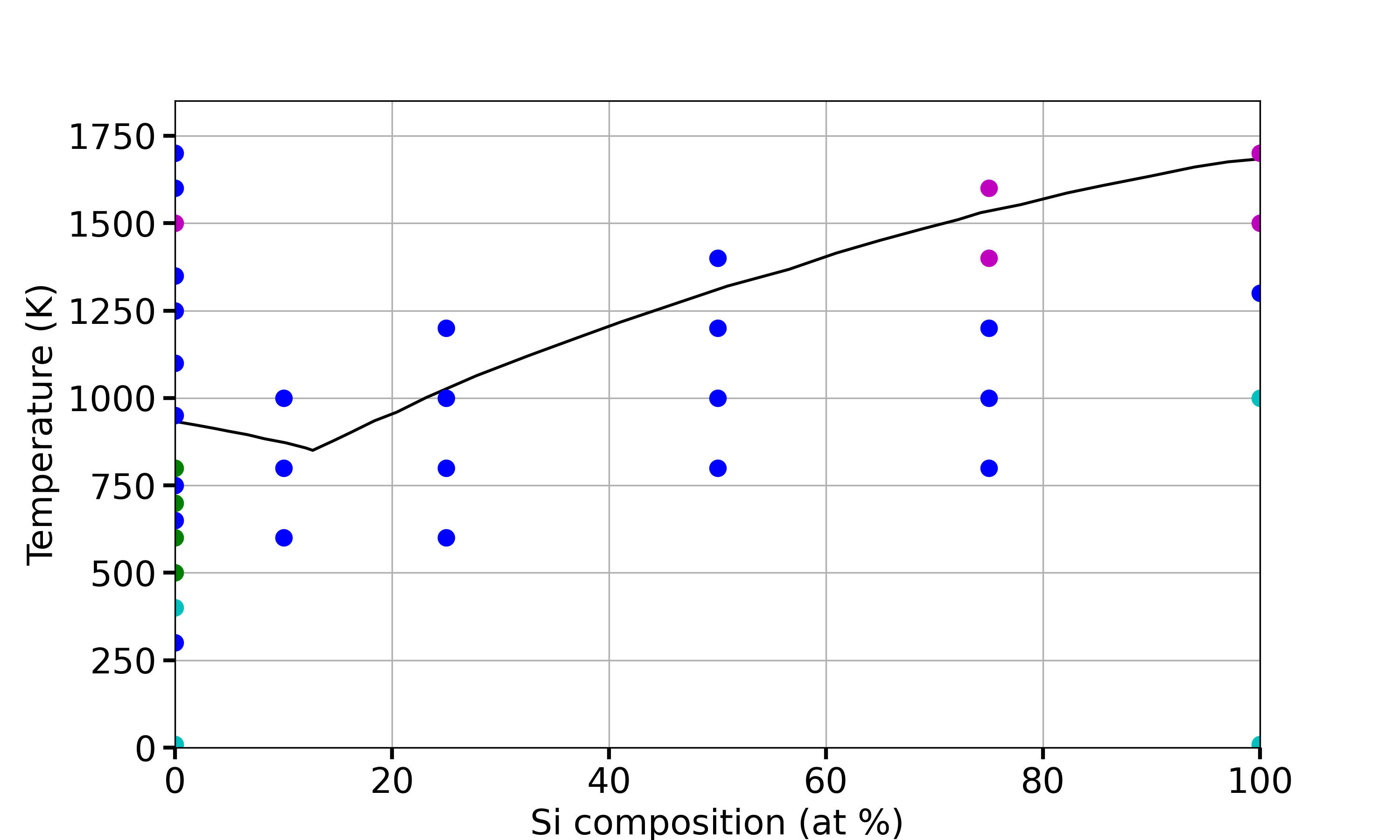} 
    \caption{Simulation map showing temperature as a function of Si composition in atomic percent. Circles indicate simulations at specific temperatures and compositions: blue for the liquid and undercooled states, cyan for crystalline states, green for crystalline and liquid states for pure Al, and magenta for high-pressure liquid simulations. The black line marks the experimental liquidus line of the Al-Si phase diagram.} 
    \label{fig:Phasediagram} 
\end{figure}

Ab initio Molecular Dynamics simulations were performed using the Vienna Ab Initio Simulation Package (VASP) \cite{kresse_ab_1993} for the purpose of generating a dataset on which the machine learning potential can be trained. 
Additionally,  independent simulations are carried out at various compositions and temperatures from which the structural and dynamics properties, as described below, are calculated to serve as a ground truth for the validation of the potential. 

The electronic behavior of the system was treated with electron-ion interactions represented by the projector augmented wave (PAW) potentials \cite{kresse_ultrasoft_1999, blochl_projector_1994}. 
The exchange and correlation effects were addressed by using the Local Density Approximation (LDA) \cite{ceperley_ground_1980}. 
This choice was made on the basis of previous studies demonstrating a good representation of the various properties for pure aluminium, especially the diffusion, but also for some aluminium alloys \cite{pasturel_chemically_2017, jakse_liquid_2013, bryk_search_2018}. 
Periodic boundary conditions were applied in the three directions of the space on a cubic cell containing $N=256$ atoms. 
As the number of atoms is sufficiently large, Brillouin zone was sampled with the $\Gamma$ point only.

The dynamical simulation were conducted by solving numerically Newton's equations of motions with the Velocity-Verlet algorithm in the velocity form using a timestep of 1.5 fs. 
The phase space trajectories were built in the canonical ensemble, \textit{i.e.} with a constant number of atoms $N$, volume $V$, and temperature $T$ (NVT) by means of the Nose-Hoover thermostat \cite{nose, hoover_canonical_1985}. Simulation volumes are adjusted so that the pressure oscillates around zero. 
In order to best reproduce short-range parts of the interactions in the MLIP, some simulations were performed with smaller volumes at high pressures as described in the supplementary information (SI) file.

The construction of our dataset was based on a mapping of the phase diagram of the binary system (Al-Si) as shown in \ref{fig:Phasediagram}.
A deep eutectic is present around $x_{Si}=0.12$ and there are no alloy crystal phases in the entire Al-Si system. Stable crystalline states were included for pure Al and Si. The data for pure Al are taken from our previous work \cite{Jakse2023}.
Together with the pure elements, compositions in the whole range were considered, namely $x_{Si}=0.10$ (close to the eutectic), $x_{Si}=0.25$, $0.50$, and $0.75$.
Several temperatures for each compositions were selected on the basis of the experimental phase diagram, considering the highest temperature slightly above the liquidus line for the each composition. All the simulated state points are indicated in Fig. \ref{fig:Phasediagram}. 
%For example, for the $x_{Si}=0.50$ composition, the highest simulated temperature is 1400K (just above 1300K in the phase diagram). 
%Then, canonical simulations, decreasing in 200K steps meaning that we have essentially undercooled liquids. For example, for the $x_{Si}=0.50$ composition: 1200K, 1000K, and 800K. 
For pure Si, the same strategy as for alloys was followed, adding diamond crystalline state for temperatures of $10$~K and $1000$~K. 
%As a database on pure aluminum had already been created in our team, we included the various simulations in our dataset. Note that simulations in the fcc solid phase are also added for pure aluminum.  The solid-liquid line is also drawn to show where our simulations lie in relation to the expected phase diagram.
The conditions under which our AIMD simulations were carried out are described in the next section, together with a description of the dataset used to build our potential.

\subsection{High-Dimensional Neural Network Potential}

%The  requires trajectories given by \textit{ab initio} molecular dynamics simulations. 
An essential aspects of the construction of a machine learning interatomic potential is to consider a relevant composition and temperature range so that the potential probes the configuration space dedicated to the study under consideration. 
%Then the details of our potential parameters will be described. To validate our approach, a comparison between our AIMD simulations, the constructed HDNNP potential and experiments will be made on structural and dynamic properties.

Machine learning interatomic potentials such as HDNNP framework are based on a nearsightedness principle for which the total energy $E_{\text{total}}$ of $N$ atoms is the sum of atomic energies, namely 
\begin{equation}
E_{\text{total}} = \sum_{i=1}^{N} E_i,
\end{equation}
where $E_{i}$ is the local energy of atom $i$. In the framework of neural networks it is exploited as follows
\begin{equation}
E_i = f_\theta(D_i),
\end{equation}
with $f_{\theta}$, a function depending on the set of parameters $\theta$ of the neural network architecture. 
The descriptor $D_{i}$ of the central atom $i$ is a feature vector representing its local atomic environment. 
In the present work, the Behler-Parrinello atom centered symmetry functions (ACSF) \cite{behler2011atom} were used. 
They are comprised of radial or angular terms describing respectively the distance and angle distributions between the atom $i$ and its surrounding neighbors, and possess the transnational and rotational invariance by definition. 
The following radial function
\begin{equation}
G_i^2 = \sum_{j \neq i} e^{-\eta (R_{ij} - R_s)^2} f_c(R_{ij})
\end{equation}
and angular functions 
\begin{equation}
\begin{split}
G_i^5 = 2^{1-\zeta} \sum_{j \neq i} \sum_{k \neq i, j} \left( 1 + \lambda \cos \theta_{ijk} \right)^\zeta  e^{-\eta \left( R_{ij}^2 + R_{ik}^2 + R_{jk}^2 \right)} \\
 f_c(R_{ij})  f_c(R_{ik})  f_c(R_{jk})
\end{split}
\end{equation}
were considered in the present work. 
$R_{ij}$ represents distance between central atom $i$ and atom $j$, and $\theta_{ijk}$ represents the angle between the atom $i$ and two neighboring atoms $j$ and $k$. 
$f_c(R_{ij})$, $f_c(R_{ik})$ and $f_c(R_{jk})$ are cutoff functions beyond which they tend smoothly to zero and interactions between atoms are neglected, thus defining the local environment extension. 
The parameters $\eta$, $\zeta$, $\lambda$, and $R_s$ are given a set values to cover the radial and angular information of the local atomic environment in the Al-Si binary alloy. These values as well as the hyper-parameters can be found on \textit{Materials Cloud} \cite{bizot_crystal_2025}.

Simulations at specific compositions and temperatures are provided in the table included in the supplementary file, and the dataset we used is also available on \textit{Materials Cloud} \cite{bizot_crystal_2025}. 
To construct this dataset, a number of 750 configurations were extracted from each simulation.
The dataset was then split into two subsets: 90 percent for training and 10 percent for testing. 
The test dataset was subsequently used to evaluate model performance, particularly through RMSE. 
The learning curve with RMSE values per atom (including those from the training set) are drawn in the supplementary file, where we selected the final potential corresponding to the minimum of this curve. 
A comparison between AIMD reference values and HDNNP test data is also provided in the supplementary file, showing excellent agreement across all configurations.
Note that we have exclusively considered energy in our training to maintain consistency with our previous works. This approach has proven effective in developing potentials for studying nucleation phenomena, which is the focus of this study.

\subsection{Classical molecular dynamics simulations}

In this work, all simulations were performed with the large-scale massively parallel atomic/molecular simulator (LAMMPS) software using the \texttt{ml-hdnnp} package implemented in 2019 \cite{singraber_library-based_2019}. 
The post-analysis is performed using OVITO software \cite{stukowski_visualization_2009} for visualization purpose, and calculation of pair-correlation, and identification of local atomic structures \textit{via} the Polyhedral Template Matching (PTM) method \cite{larsen_robust_2016}. 
Complementary analyses, including mean-square displacement (MSD), pair-correlation functions, and structure factors, were carried out with the ISAACS software \cite{le_roux_isaacs_2010}.
In addition, simulation data management, such as transformation of position file produced by VASP into an input file for Lammp, as well as extraction and sampling of the simulated trajectories were done with the Atomic Simulation Environment (ASE) code \cite{larsen_atomic_2017}.

Although the RMSE on energy gives a good indication of the performance of a MLIP training, it is necessary to characterize the structural and dynamic quantities with the trained HDNNP potential and compare them with the AIMD results to validate it. 
Ultimately, validation with respect to existing experimental data provides the reliability of the MLIP as well as \textit{ab initio} simulations.
To this end, our MD simulations are carried out under the same conditions as the AIMD, \textit{i.e}. in the NVT (constant number of atoms, $N$, constant volume, $V$, and constant temperature, $T$) ensemble with a Nose-Hoover thermostat. 
For homogeneous nucleation simulations, the NPT ensemble (constant pressure $P$ is chosen to allow the volume to adjust appropriately during the liquid-solid phase transition.

\begin{figure*}[t] 
    \centering
    \includegraphics[angle=0, width=0.9\textwidth]{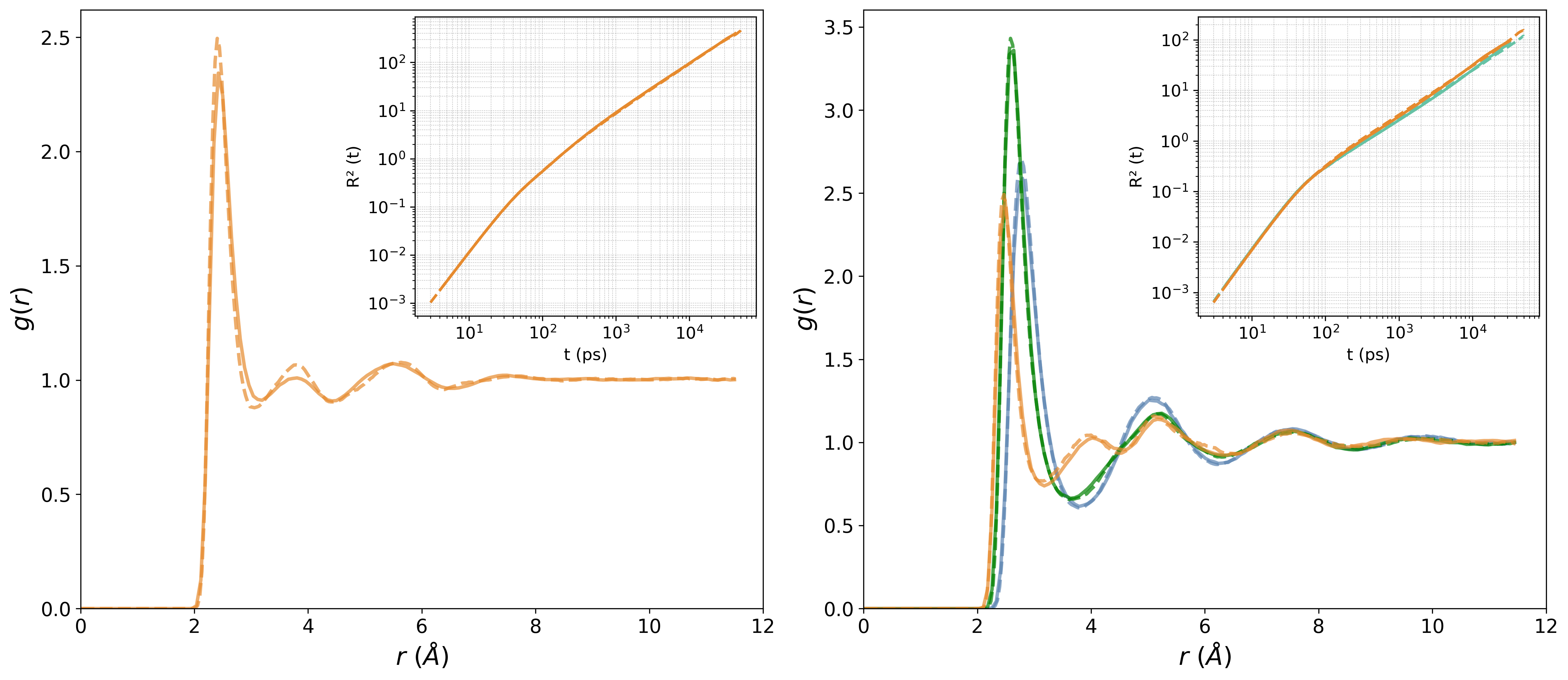} 
    \caption{Comparison of Mean Square Displacement $R^2(t)$ and radial distribution functions, $g(r)$ between HDNNP and AIMD for $x_{Si}=$0.5 at $800$ K (right), and pure Si at $1300$ K (left). Dashed lines represent AIMD simulations, and solid lines are MD simulations with the HDNNP. Orange, green, and blue colors represent Si-Si, Al-Al, and Al-Si pairs for radial distribution functions. Orange and green colors represent Si atoms and Al atoms for Mean Square Displacement. } 
    \label{fig:RDFMSD}
\end{figure*}

\subsection{Structural and dynamic properties}

The partial pair-correlation function $g_{ij}(r)$, indicates the probability that an atom of type $i$ will be found at a distance $r$ from an atom of type $j$. 
This probability can be estimated in simulations by quantifying the number of $j$ atoms within a spherical shell of radius $r$ and width $\Delta r$ given by
\begin{equation}
g_{ij}(r) = \frac{N_j}{V} \frac{n_{ij}(r)}{4\pi r^2 \Delta r}
\end{equation}
$n_{ij}(r)$ represents the average, $V$ the volume of the simulation box and ${N_j}$ the number of $j$ atoms. The Fourier transform of the partial pair-correlation gives the partials structures factors written in the Faber-Ziman form using the Debye formula:
\begin{equation}
S_{ij}(q) = \frac{1}{N} \left\langle \sum_{k=1}^{N_i} \sum_{l=1}^{N_j} \exp \left( i \mathbf{q} \cdot (\mathbf{r}_k^i - \mathbf{r}_l^j) \right) \right\rangle
\end{equation}
The total structure factor $S(q)$ is a weighted sum of the partial structure factors, reflecting the contributions of atom pairs between each type of species.

In order to evaluate the self-diffusion coefficients, the Mean-Square Displacement (MSD) as a function of time $t$ is considered. It reads
\begin{equation}
R_i^2(t) = \frac{1}{N_i} \sum_{j=1}^{N_i} \left\langle [r_j(t + t_0) - r_j(t_0)]^2 \right\rangle,
\end{equation}
where the summation is over atoms of species $i$, $r_{j}(t)$ represents the position of atom $j$ of species $i$ at time $t$, and $<.>$ is the average taken over sufficient number of initial times $t_{0}$. 
The diffusion coefficient of each species $i$, $D_i$, is finally extracted from the slope of the MSD curve in its linear part at long times, namely
\begin{equation}
D_i = \lim_{t \to \infty} \frac{R_i^2(t)}{6t},
\end{equation}
where the factor $6$ refers to the diffusion in the three-dimensional space.

\section{Results}

\label{sec:results}

\subsection{Validation}

As the present machine learning interatomic potential is mainly dedicated to the crystal nucleation, it is important that the properties of the liquid, such as the local structural and dynamic properties are well reproduced with respect to AIMD simulations. For the sake of consistency, MD simulations with $N=256$ atoms at the same volume as for the AIMD were performed in the NVT ensemble with a Nosé-Hoover thermostat. 
Equilibration was first carried out at the temperature of interest during $20$ ps, then runs were continued further for $50$ ps during which the targeted quantities were produced and averaged. 
A range of Si compositions ($x_{Si}$) and temperatures were explored, selecting the following thermodynamic states $x_{Si}=0.5$ at $800$ K, and $1$ at $1300$ K in Fig. \ref{fig:RDFMSD}. 
Results for other compositions and temperatures are given in supplementary material file. 
The RDFs simulated with the MLIP are in excellent agreement with AIMD data. 
Note that peak intensity for pure Si is slightly lower for our potential, but with peaks positions very well reproduced. 
As for the MSDs, they are extremely well reproduced by the potential, with remarkable superposition for Si and for all the alloy compositions, as exemplified for $x_{Si} = 0.5$ in Fig. \ref{fig:RDFMSD}.

\begin{figure} 
    \includegraphics[angle=0, width=0.45\textwidth]{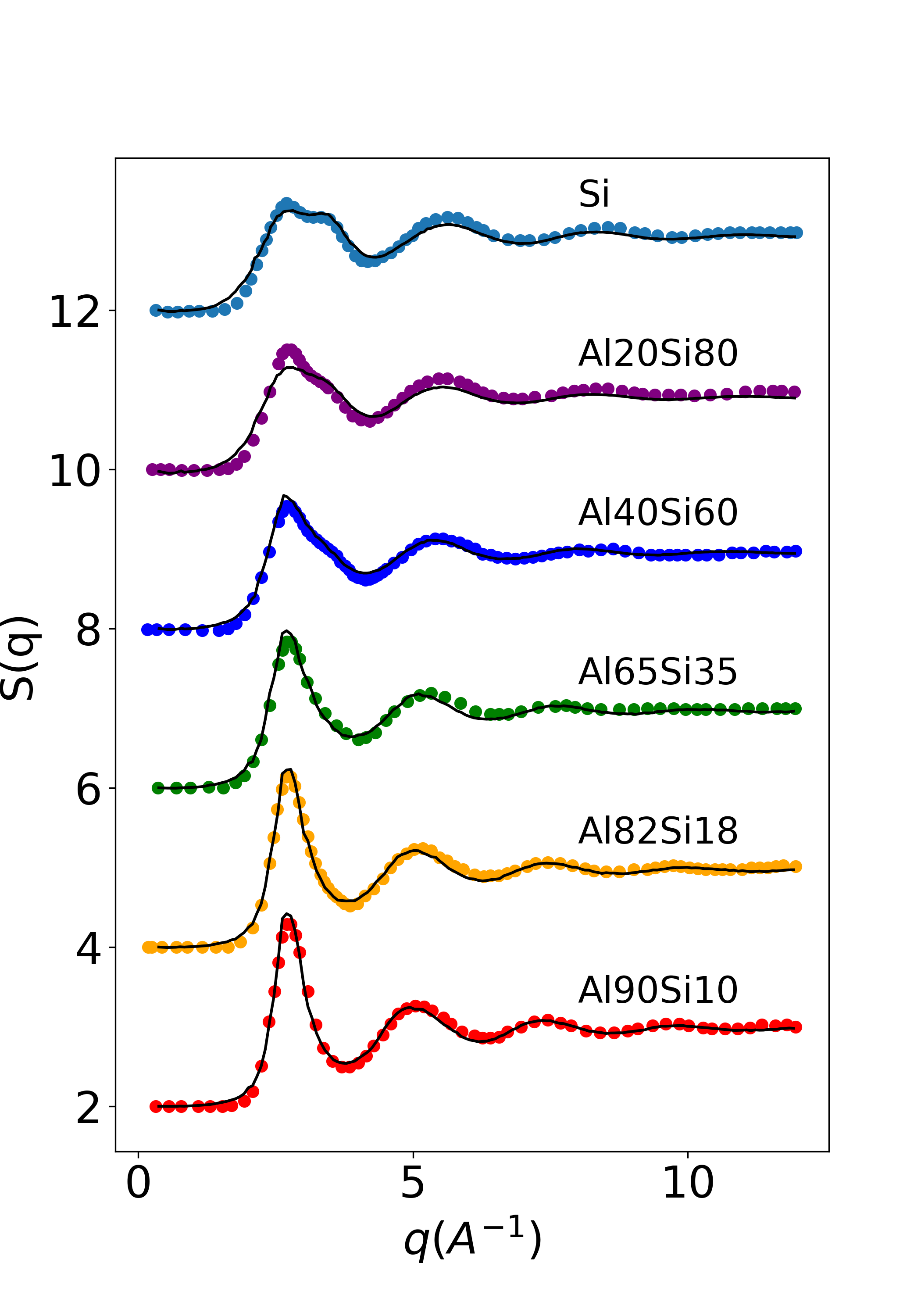} 
    \caption{Comparison of the total structure factors between the experiments (circles, \cite{kazimirov_x-ray_2013}) and the ML potential (black solid line). The composition with 10, 18, 35, 60, 80, and 100 percent are represented with in red, orange, green, and blue respectively. } 
    \label{fig:SF} 
\end{figure}

Regarding experimental data in the literature for the Al-Si system in the liquid phase, the difficulty of measurements lies mainly in the very weak contrast difference between aluminum and silicon. 
In addition, low solubility complicates the analysis of liquid alloys by introducing heterogeneous atomic structures, making it very difficult to measure properties such as viscosity, diffusion coefficients or density.
This is why most of the dynamic properties calculated for this alloy are performed by Molecular Dynamics \textit{via} classical potentials or by \textit{ab initio} calculations, and most often for compositions close to the eutectic (around $x_{Si}$=0.12). 
Nevertheless, total structure factors in the liquid phase were measured over the entire composition range \cite{kazimirov_x-ray_2013}, providing us a reference for the local structure. 
Comparison with the results of our MLIP are presented in Fig. \ref{fig:SF} for compositions $x_{Si} = 0.1$, $0.18$, $0.35$, $0.6$, $0.8$ and $1$, at respective temperatures of $893$ K, $1003$ K, $1223$ K, $1483$ K, $1653$ K, and $1743$ K.
As the experimental densities are not given in the literature, to the best of our knowledge, at these temperatures and compositions specifically, MD simulations with the HDNNP were carried out simulations in the NPT ensemble at ambient pressure. Equilibration was performed for $50$ ps and quantity were produced subsequently for $250$ ps.
Total structure factors were then calculated at the equilibrated average volume of these simulations in the production part using NVT conditions. They are drawn in Fig. \ref{fig:SF} and show a good agreement with experiments. 
The peaks are well located, and their amplitudes are also well described, especially in the rich Al region near the eutectic. 
The progressive broadening and splitting of the first peak towards the Si-rich part ($x_{Si}=$0.8 and 1) is observed for the MLIP, although the amplitude is slightly lower than experiments. 
This feature is linked to the fact that Si keeps partially covalent bonds in the liquid, broadening the distribution of interatomic distances, and leading to broader peaks in the structure factors. 
By contrast, the Al-rich part is characterized by metallic bonding leading to a more compact structure.
From these results, our potential is seen to be successful in describing both types of bond over the entire composition range. 

\subsection{Crystal nucleation in Al-Si}

\begin{figure*}[t] 
    \includegraphics[angle=0, width=0.95\textwidth]{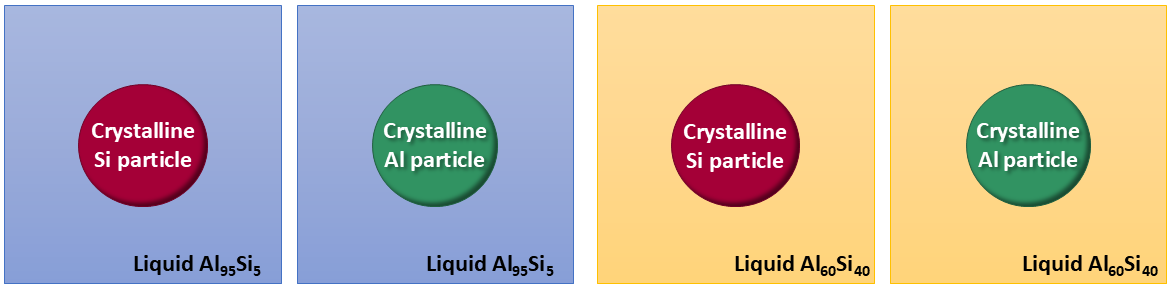} 
    \caption{Schematic view of the initial geometry of the simulations. On the left are represented the two cases in the hypoeutectic composition with solid Al and Si particles. On the right are represented the two cases in the hypereutectic composition with solid Al and Si particles.} 
    \label{fig:Schematic} 
\end{figure*}

To observe nucleation in the Al-Si eutectic alloy, a specific simulation setup was designed as sketched in Fig. \ref{fig:Schematic}. 
First, a super-cell consisting solely of aluminum in the fcc phase was created with a lattice parameter of $4.05$ Å at $300$ K, containing a total of $108000$ atoms. 
To melt the aluminum, the temperature was gradually increased from $300$ K to $1300$ K at ambiant pressure using a linear ramp, ensuring full melting. 
The system was then cooled down to $600$ K to obtain a undercooled liquid state with a cooling rate of $10^{12}$ K/s.
finally, to create an alloy in the hypo-eutectic region, $5$ percent of Al atoms in the liquid phase were randomly replaced with Si atoms and the system was equilibrated again at this temperature.
In the second step, a crystalline Si or Al particle of approximately $7 100$ atoms, maintaining their stable crystalline structures (\textit{i.e.} diamond cubic for Si and fcc for Al), at the center of the simulation box. 
 Similarly, for the hyper-eutectic region, we replaced $40$ percent of Al atoms with Si atoms.

To ensure the stability of the system, a thermalization was applied in the NPT ensemble at $600$ K after the insertion of the crystalline particle. 
This setup allowed us to investigate the nucleation process in both hypo- and hyper-eutectic compositions of the Al-Si alloy. 
Thus, four different cases were considered here: two in the hypo-eutectic conditions, one with a solid Si particle (Fig. \ref{fig:Schematic}, left) and one with a solid Al particle (Fig. \ref{fig:Schematic}, middle left), and two in the hyper-eutectic composition (Fig. \ref{fig:Schematic}, middle right and right), with exactly the same particles sizes.

\begin{figure*}[t] 
    \includegraphics[angle=0, width=0.95\textwidth]{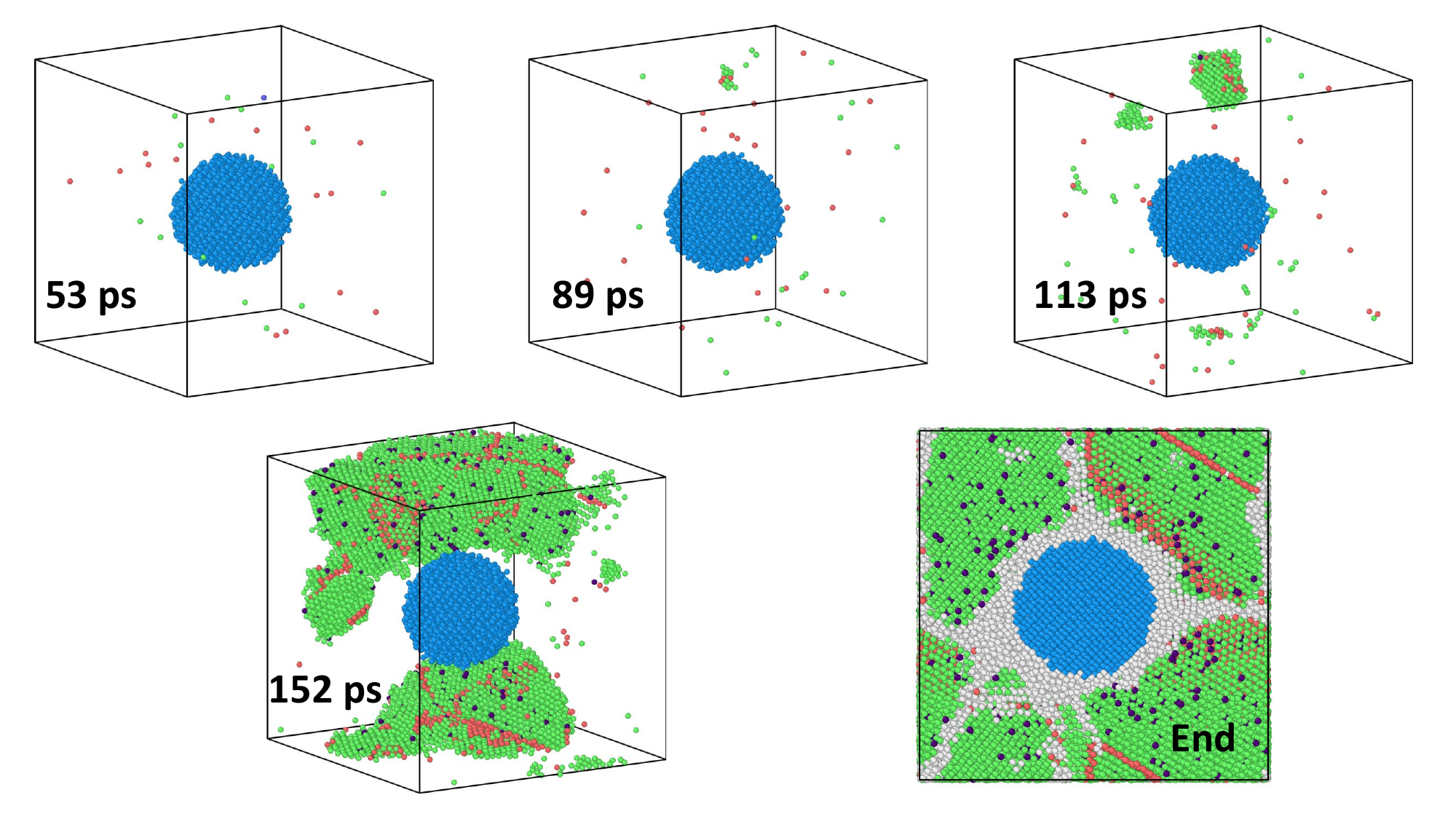} 
    \caption{Snapshots of the evolution of the simulation with respect to the time at $53$, $89$, $113$, $152$ ps and the final microstructure. Dark blue, red, and green colors represent hcp or fcc Si, hcp, and fcc local structures, respectively, as identified with the PTM approach (see text). Light blue color represents the Si atoms in a diamond local structure. Atoms in the unknown structure in the sense of the PTM are considered as liquid and are not drawn to give a better representation of the nucleation pathway. In the latest frame for the solidified configuration, unknown structures that remain are drawn in the grey color.} 
    \label{fig:Evolution} 
\end{figure*}

%In contrast to the experiments, the nucleation temperature at atomic scales needs to be relatively large. For this reason, nuclei grow at velocity higher than the liquid Si diffusion and end up systematically capturing Si atoms. However, at slower growth rates, Si atoms are rejected at the periphery, probably indicating that, at lower nucleation temperatures, we would observe nuclei composed solely of aluminium atoms.
\begin{figure*}[t] 
    \includegraphics[angle=0, width=0.8\textwidth]{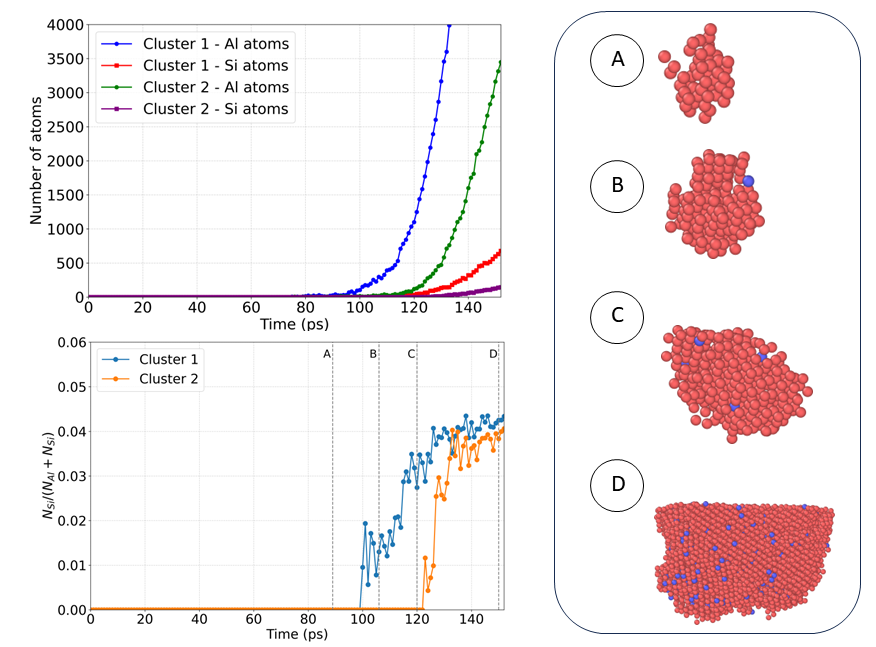} 
    \caption{Evolution of the number of atoms in nuclei 1 and 2 over time. The top left panel shows the number of Al and Si atoms in these nuclei as a function of time, while the bottom left panel presents the ratio of Si atoms within the nuclei over time. On the right, four snapshots capture the main nucleus at 89 ps (A), 103 ps (B), 120 ps (C), and 150 ps (D).} 
    \label{fig:EvolutionCluster} 
\end{figure*}

\subsubsection{Hypo-eutectic nucleation : case of Si particle}

During the thermalization at $600$ K, no nucleation was observed within the simulated time window, the degree of undercooling being too small. 
The temperature was further decreased from $600$ K to $550$ K in $100$ ps, and maintained at a constant temperature of $550$ K until the entire system is solidified. 
The global evolution of the simulation is depicted in Fig. \ref{fig:Evolution}. 
up to $89$ ps for the present simulation, pre-critical nuclei appear and dissolve back in the liquid phase. Then a first critical nucleus is observed in the liquid phase, while the Si particle remains crystalline in its diamond phase in the surrounding liquid with a only slight growth during the time of the simulation. 
%The temperature at this stage is around 575 K. 
Atoms giving rise to nucleation at this stage are only aluminum ones, and no Si atoms are present in the nucleus. 
From $89$ ps to $113$ ps, the first nucleus continues to grow, with a majority of aluminum atoms with solute trapping, namely a few Si atoms captured due to the very rapid growth of the nucleus. 
Around $113$ ps, another nucleus appears in the liquid phase, with the same nucleation mechanism as the first nucleus. Note that fcc/hcp stacking faults are observed across the grains formed. 
This behavior is well described in the literature for Al solidification \cite{mahata_understanding_2018}. 
From 152 ps to the complete solidification, other nuclei appears and grow similarly. 
The final \textit{microstructure} shows us that no nucleation arises from the interface of the Si particle. 
Grain boundaries are observed when 2 adjacent grains impinge during growth. A zone around the Si particle is observed with a higher Si content.
As a matter of fact, the amount of Si in the grain boundaries is estimated to be around $10$ percent, higher than the chosen liquid composition at the initial stage before nucleation, indicating obviously that Si atoms are rejected during growth.

To quantitatively identify the atomic composition of the clusters, a focus is made on the first two clusters that appear in the simulation box. 
Since the underlying mechanisms for subsequent clusters are identical, they are excluded from the analysis for the sake of clarity.
Fig. \ref{fig:EvolutionCluster} illustrates the evolution of the number of atoms in nuclei 1 and 2 over time. The top left panel shows the number of Al and Si atoms in these nuclei as a function of time, while the bottom left panel presents the ratio of Si atoms within the nuclei over time. On the right hand side, four snapshots capture the system at $89$ ps (A), $103$ ps (B), $120$ ps (C), and $150$ ps (D). Focusing specifically on cluster 1, the evolution of the simulation, can be divided into several distinct regimes. 
At 89 ps, the first nucleus appears with a very slight increase in the number of Al atoms. However, the number of Si atoms remains unchanged. This is confirmed by the figure at the bottom left, which shows the fraction of Si atoms as a function of time. At the time marked by (A), the number of Si atoms in the nucleus is zero.
Between 89 ps and 113 ps, the number of Al atoms in the nucleus increases exponentially. During this stage, a slight rise in the number of Si atoms is also observed (Fig. \ref{fig:EvolutionCluster}, bottom left). However, as seen in snapshot B, this increase is minimal at 103 ps, only a single Si atom is present in the cluster, positioned at the border.
From $120$ ps, the number of Al atoms in the cluster increase further, initiating a linear growth. At this stage, a more noticeable increase in Si atoms is also observed.
Snapshot C reveals that from this point, Si atoms begin to be incorporated into the nucleus, suggesting that nucleus growth outpaces Si diffusion in the liquid phase. 
Finally, at $150$ ps, both Al and Si nuclei continue to grow. The number of Si atoms reaches a saturation level of approximately 4 percent, close to the initial liquid composition, where they become fully incorporated into the fcc-Al matrix (snapshot D). 
For this case, the evolution of nucleation demonstrates that nucleation is homogeneous and occurs only in regions where there are exclusively Al atoms and sufficient space (i.e without Si atoms) to exceed the critical radius.

\subsubsection{Hypo-eutectic nucleation : case of solid-Al particle}

\begin{figure}[t] 
    \includegraphics[angle=0, width=0.45\textwidth]{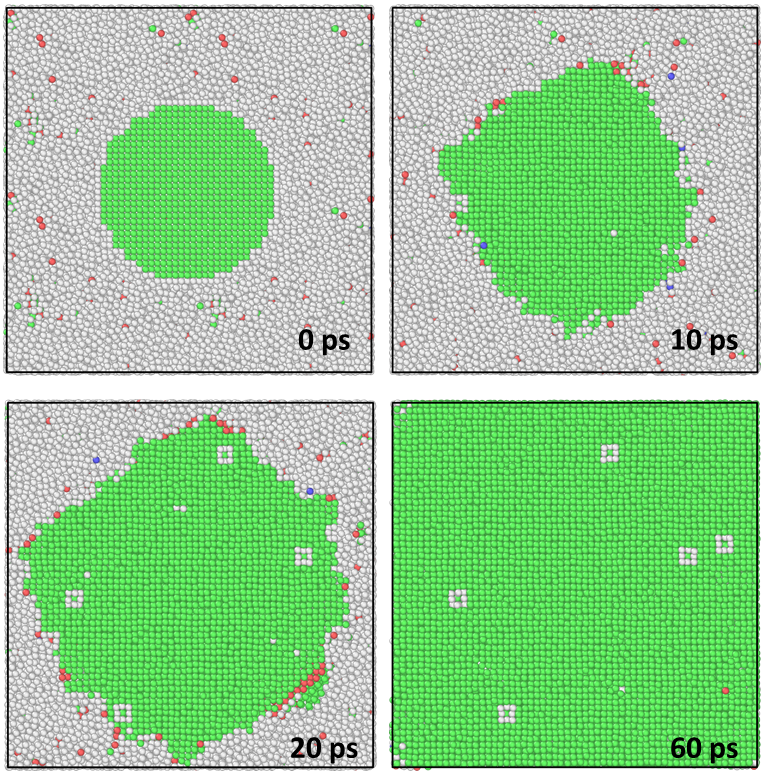} 
    \caption{Snapshots at time 0 ps, 10 ps, 20 ps, and 60 ps of the evolution of the simulation in the hypoeutectic composition in presence of the solid Al particle. Green, red, and white represents fcc, hcp, and unknown structure.} 
    \label{fig:LowcompPartAl} 
\end{figure}

Figure \ref{fig:LowcompPartAl} illustrates the evolution of the simulation, where an Al seed is included. At 0 ps, the Al particle remains spherical at the center of the box, in the fcc structure and surrounded by liquid atoms. Solidification begins immediately at the surface of the Al seed, and within $10$ ps, an increase in fcc atoms is visible, especially at the top and left of the particle. At 20 ps, the particle expands preferentially along the [100], [010] and [001] directions, which is expected for fcc-Al due to its fastest growth rate in this orientation. By 60 ps, the entire system has fully solidified in the fcc structure.

As in the previous case, solidification is rapid enough to outpace Si diffusion in the liquid phase, leading to the instantaneous trapping of Si atoms into the fcc-Al matrix.

\subsubsection{Hyper-eutectic nucleation : case of solid Si and Al particle}

\begin{figure}[t] 
    \includegraphics[angle=0, width=0.45\textwidth]{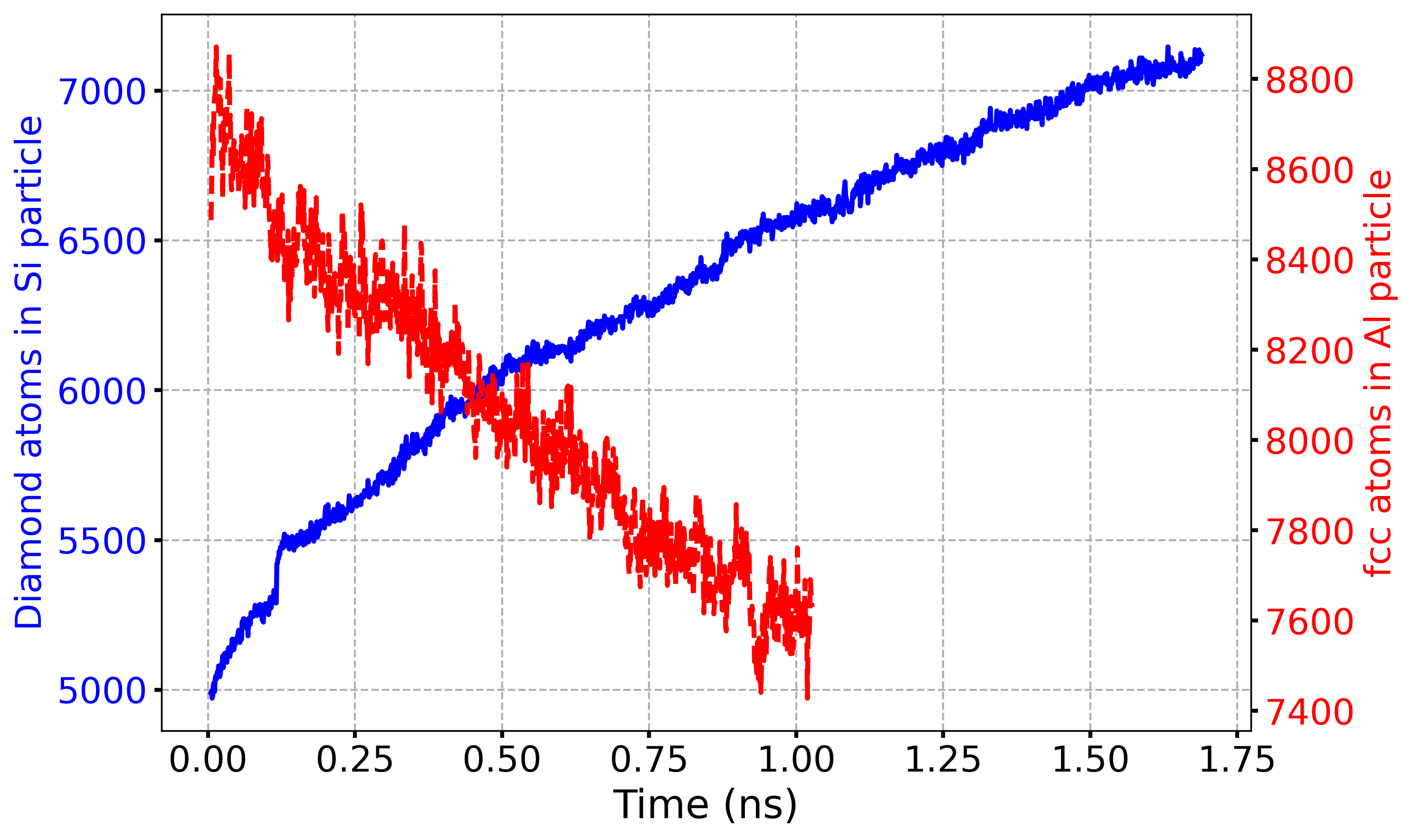} 
    \caption{Blue : Evolution of the number of Si atoms in diamond structure (i.e. the solid Si particle) in function of the time. Red : Evolution of the number of Al atoms in fcc structure (i.e. the solid Al particle)  in function of the time.} 
    \label{fig:AlSinumber} 
\end{figure}

For the Si seed in the hyper-eutectic composition, no nucleation at all occurs during the time span of the simulation. 
This can be explained by the fact that, due to the higher concentration of Si conditions where Al atoms can form an embryo and and grow to exceed the critical radius is rare. This is same for Si embryos, and we observed a slow increase in the number of Si atoms in the diamond structure over time as shown in Fig. \ref{fig:AlSinumber}. The particle gradually grows to around 100 atoms per 100 ps. As the particle grows, it also becomes highly faceted in a polygonal shape. 

In contrast, the solid Al seed follows an opposite trend. Over time, we observe a gradual decrease in its size as shown in Fig. \ref{fig:AlSinumber}, suggesting partial dissolution into the surrounding liquid. 
In addition, although the overall trend is the decrease, the oscillations are more pronounced than for the Si particle due to density and concentration fluctuations in the surrounding liquid. 
Increasing, the temperature closer to the liquidus line speeds up this nucleation mechanism without changing it.
The global decay is attributed to Si atoms, higher in concentration in the liquid surrounding the particle, which gradually destabilize the solid Al seed. 
Unlike the Si seed, which adopts a well-defined polygonal shape, the Al seed seems to be unstable and dissolve when the composition is in the hyper-eutectic range, indicating that the pure Si phase is most likely driving nucleation.

\section{Conclusion}
\label{sec:conclusion}

Some aspects of the early stages of crystal nucleation mechanisms in eutectic liquid Al-Si alloy were studied by means of molecular dynamics simulations. 
For this purpose, a machine learning interatomic potential by means of a high-dimensional neural network was trained on \textit{ab initio} molecular dynamics within the density functional theory. 
The potential was systematically validated on structural and dynamic properties against AIMD results. Results were was also compared to the total structure factors from x-ray measurements with experiments in the complete range of compositions, with excellent agreement especially for Al-rich compositions, close to eutectic point. It is shown to be able to capture the metallic as well as partial covalent bonding coexisting in this system. 

Given the time and length scales achievable by brute force molecular dynamics simulations considered here, a seeding scheme was designed, in which either pure Al or pure Si crystalline particles are introduced into the liquid alloy. 
Two compositions were considered, one in slight hypo-eutectic condition containing $5$ atomic percent Si, and one in hyper-eutectic conditions with $40$ atomic percent Si (hyper-eutectic). 
This approach allows us to put forward the following scenario for the early stages of the eutectic nucleation.
In slight hypo-eutectic conditions, our findings indicate that nucleation initiates only with pure aluminum embryos both with or without seeding particles. The surrounding liquid alloy composition is progressively enriched in Si, with Si atoms being only partially trapped in nuclei by rapid growth. 
In the case of seeding with Al particle, the growth was instantaneous from it, while for the seeding with Si, homogeneous nucleation of Al occurs the liquid phase.  
Additionally, in the hyper-eutectic, no rapid nucleation was observed both with Al and Si seeding. Rather, progressive dissolution of the Al particle was always seen, while the Si particle grows slowly and linearly, with polygonal faceting. Finally, for the Al-Si alloy in hypo-eutectic conditions, the dominant emerging phase seems to be aluminum first, whereas in hyper-eutectic situations Si phase drives the eutectic solidification process.

\section{Acknowledgments}

We acknowledge the CINES, TGCC and IDRIS under Project No. INP2227/72914/gen5054, as well as CIMENT/GRICAD for computational resources, as well as financial support under the French-German project PRCI ANR-DFG SOLIMAT (ANR-22-CE92-0079-01). This work has been partially supported by MIAI@Grenoble Alpes (ANR-19-P3IA-0003). Discussions within the French collaborative network in artificial intelligence in materials science GDR CNRS 2123 (IAMAT) are also acknowledged.

\bibliographystyle{apsrev}
\normalem
\bibliography{references.bib}
\end{document}

\maketitle

\begin{center}
    \textbf{Table of Contents}
\end{center}
\hrule
\vspace{0.5cm}

\tableofcontents

\newpage

\section{Simulation background}

\subsection{Dataset: Ab-Initio Molecular Dynamics trajectories}

\begin{longtable}[c]{|c|c|c|}
    \hline
    \multicolumn{3}{|c|}{\textbf{Pure Al}} \\
    \hline
    \textbf{Temperature (K)} & \textbf{Pressure (kB)} & \textbf{State} \\
    \hline
    \endfirsthead % Définition de l'en-tête qui sera répété à chaque page
    \multicolumn{3}{c}{\textbf{Table \thetable\ continued}} \\ % Texte affiché à chaque page suivante
    \endhead
    
    10    & 0  & Solid \\
    300   & 0  & Liquid \\
    400   & 0  & Solid \\
    500   & 0  & Liquid \\
    500   & 0  & Solid \\
    600   & 0  & Liquid \\
    600   & 0  & Solid \\
    650   & 0  & Liquid \\
    700   & 0  & Liquid \\
    700   & 0  & Solid \\
    750   & 0  & Liquid \\
    800   & 0  & Liquid \\
    800   & 0  & Solid \\
    950   & 0  & Liquid \\
    1100  & 0  & Liquid \\
    1250  & 0  & Liquid \\
    1350  & 0  & Liquid \\
    1500  & 0  & Liquid \\
    1500  & HP & Liquid \\
    1600  & 0  & Liquid \\
    1700  & 0  & Liquid \\
    \hline
    \multicolumn{3}{|c|}{\textbf{Al$_{90}$Si$_{10}$}} \\
    \hline
    \textbf{Temperature (K)} & \textbf{Pressure (kB)} & \textbf{State} \\
    \hline
    600   & 0  & Liquid \\
    800   & 0  & Liquid \\
    1000  & 0  & Liquid \\
    \hline
    \multicolumn{3}{|c|}{\textbf{Al$_{75}$Si$_{25}$}} \\
    \hline
    \textbf{Temperature (K)} & \textbf{Pressure (kB)} & \textbf{State} \\
    \hline
    600   & 0  & Liquid \\
    800   & 0  & Liquid \\
    1000  & 0  & Liquid \\
    1200  & 0  & Liquid \\
    \hline
    \multicolumn{3}{|c|}{\textbf{Al$_{50}$Si$_{50}$}} \\
    \hline
    \textbf{Temperature (K)} & \textbf{Pressure (kB)} & \textbf{State} \\
    \hline
    800   & 0  & Liquid \\
    1000  & 0  & Liquid \\
    1200  & 0  & Liquid \\
    1400  & 0  & Liquid \\
    \hline
    \multicolumn{3}{|c|}{\textbf{Al$_{25}$Si$_{75}$}} \\
    \hline
    \textbf{Temperature (K)} & \textbf{Pressure (kB)} & \textbf{State} \\
    \hline
    800   & 0  & Liquid \\
    1000  & 0  & Liquid \\
    1200  & 0  & Liquid \\
    1400  & 0  & Liquid \\
    1400  & HP & Liquid \\
    1600  & 0  & Liquid \\
    1600  & HP & Liquid \\
    1600  & LP & Liquid \\
    \hline
    \multicolumn{3}{|c|}{\textbf{Pure Si}} \\
    \hline
    \textbf{Temperature (K)} & \textbf{Pressure (kB)} & \textbf{State} \\
    \hline
    10    & 0  & Solid \\
    1000  & 0  & Solid \\
    1300  & 0  & Liquid \\
    1500  & 0  & Liquid \\
    1500  & 0  & Liquid \\
    1700  & 0  & Liquid \\
    1700  & HP & Liquid \\
    1700  & LP & Liquid \\
    \hline
    \caption{Thermodynamic states considered in this study for building the dataset to train the HDNNP.}
\end{longtable}

\subsection{High-Dimensional Neural Network Potential}

\begin{figure} [h]
    \centering 
    \includegraphics[angle=0, width=0.8\textwidth]{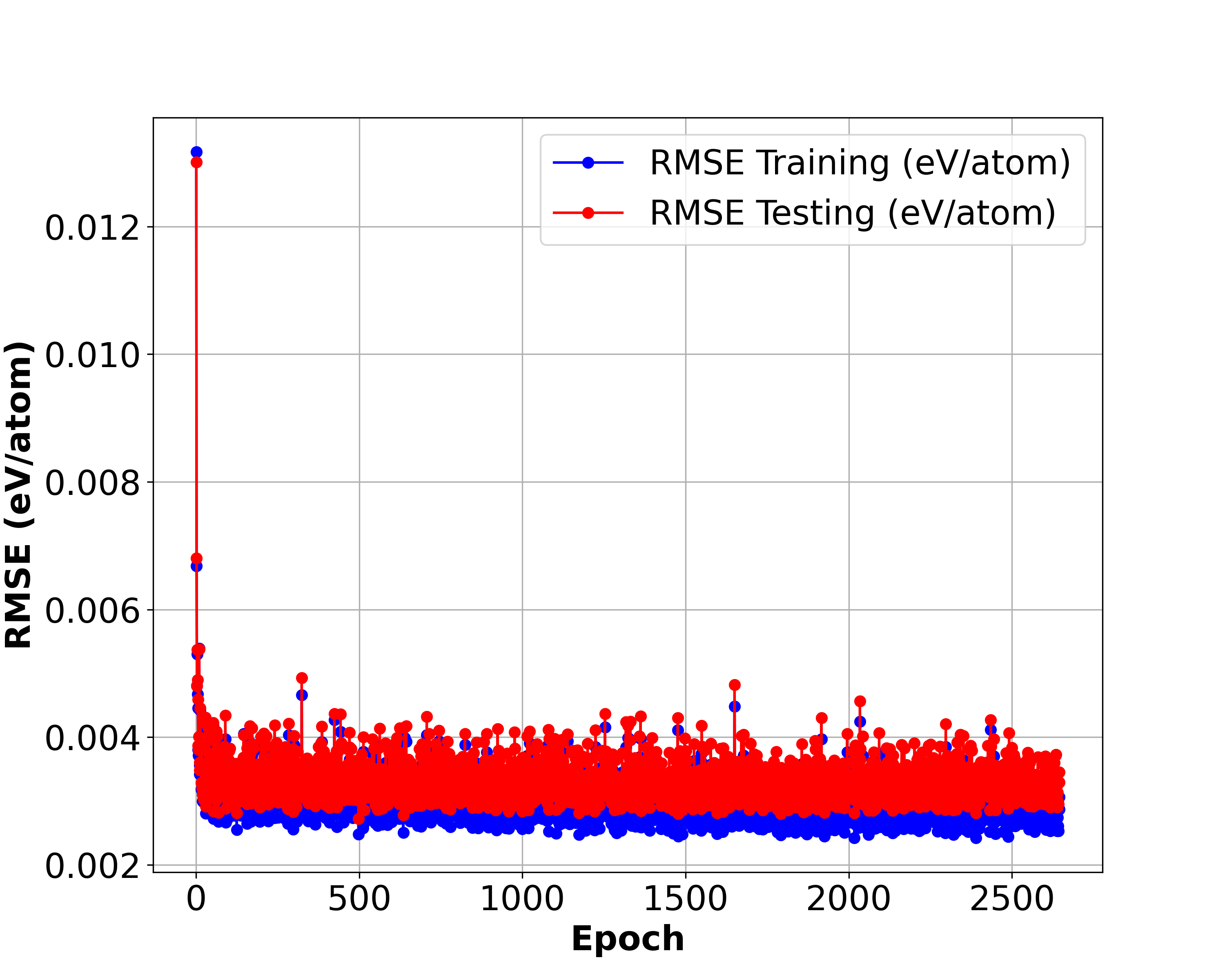} 
    \caption{RMSE in eV/atom in function of the epoch for the training and the test set.} 
    \label{fig:Phasediagram} 
\end{figure}

\begin{figure} [h]
    \centering 
    \includegraphics[angle=0, width=0.8\textwidth]{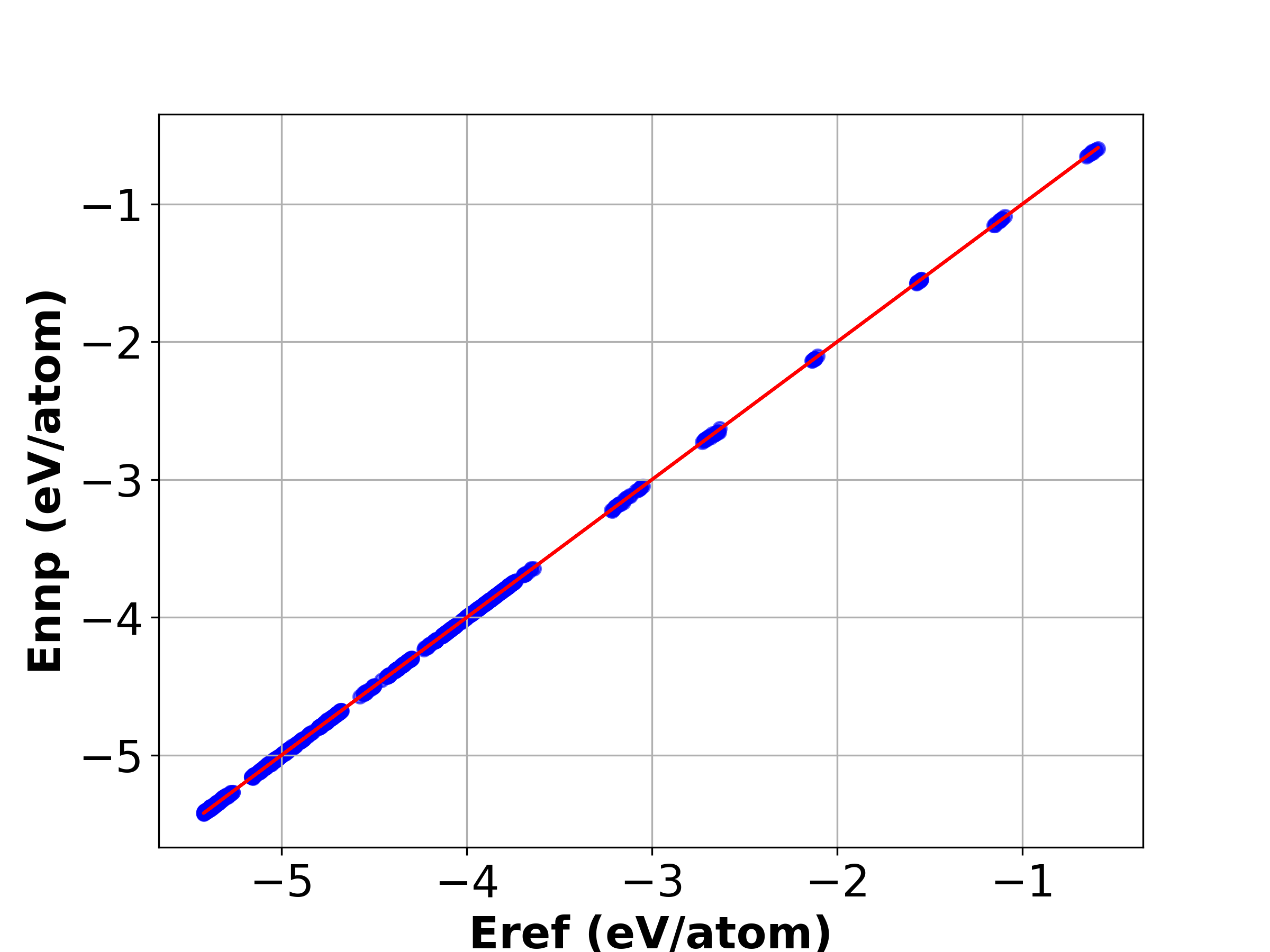} 
    \caption{AIMD (reference) energies versus HDNNP energies for the test set.} 
    \label{fig:Phasediagram} 
\end{figure}

\begin{figure} [h]
    \centering 
    \includegraphics[angle=0, width=0.8\textwidth]{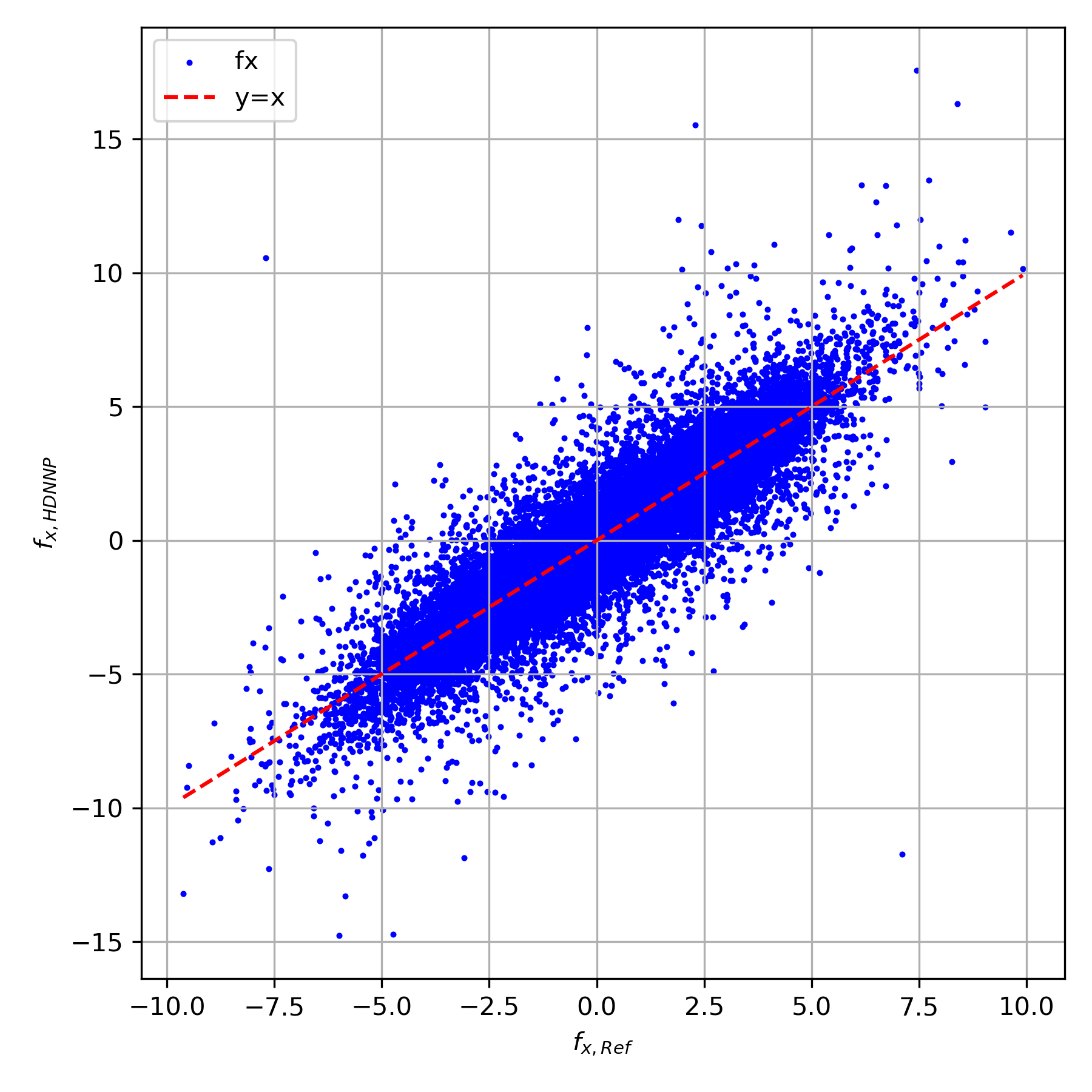} 
    \caption{AIMD (reference) x forces versus HDNNP x forces for the test set.} 
    \label{fig:Phasediagram} 
\end{figure}

\begin{figure} [h]
    \centering 
    \includegraphics[angle=0, width=0.8\textwidth]{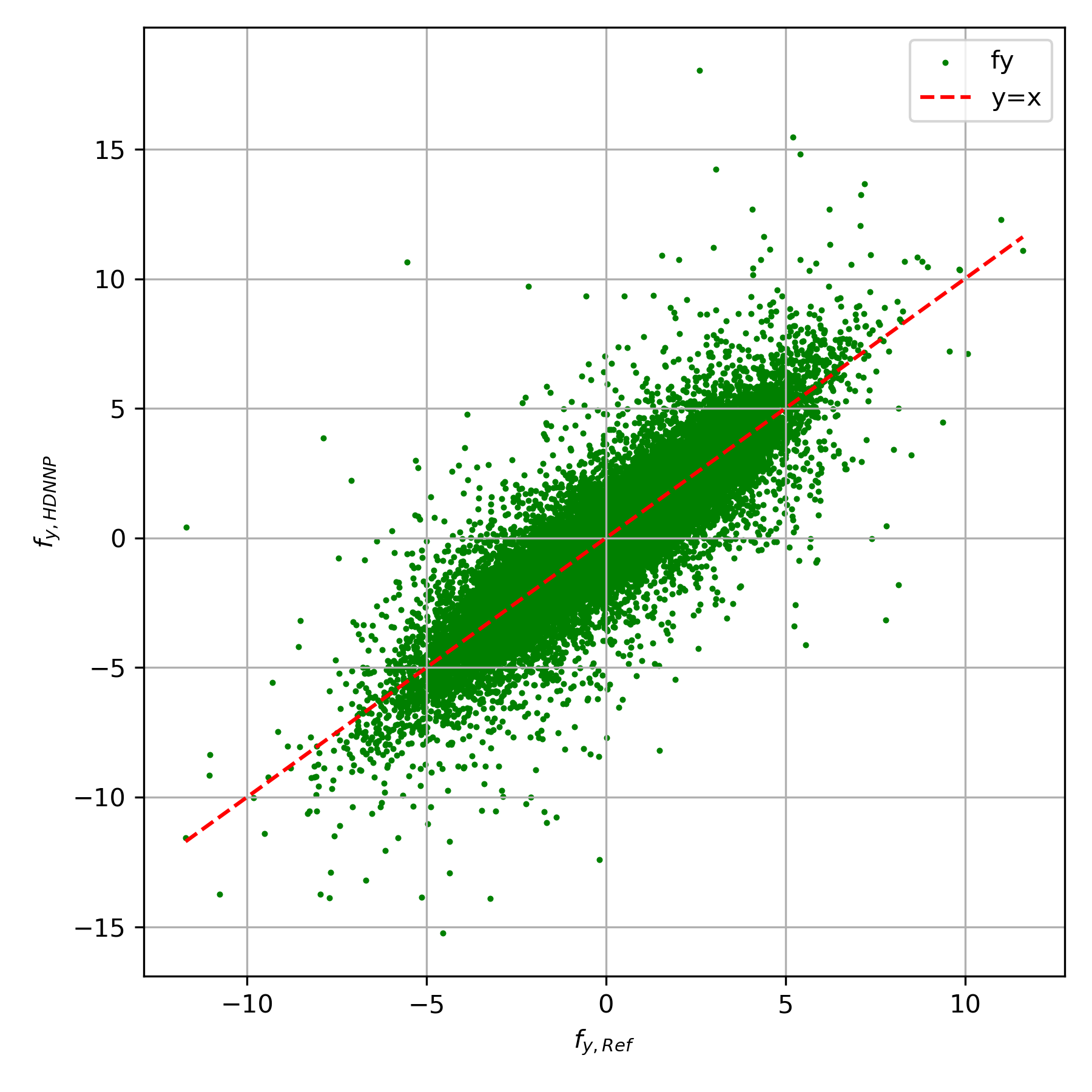} 
    \caption{AIMD (reference) y forces versus HDNNP y forces for the test set.} 
    \label{fig:Phasediagram} 
\end{figure}

\begin{figure} [h]
    \centering 
    \includegraphics[angle=0, width=0.8\textwidth]{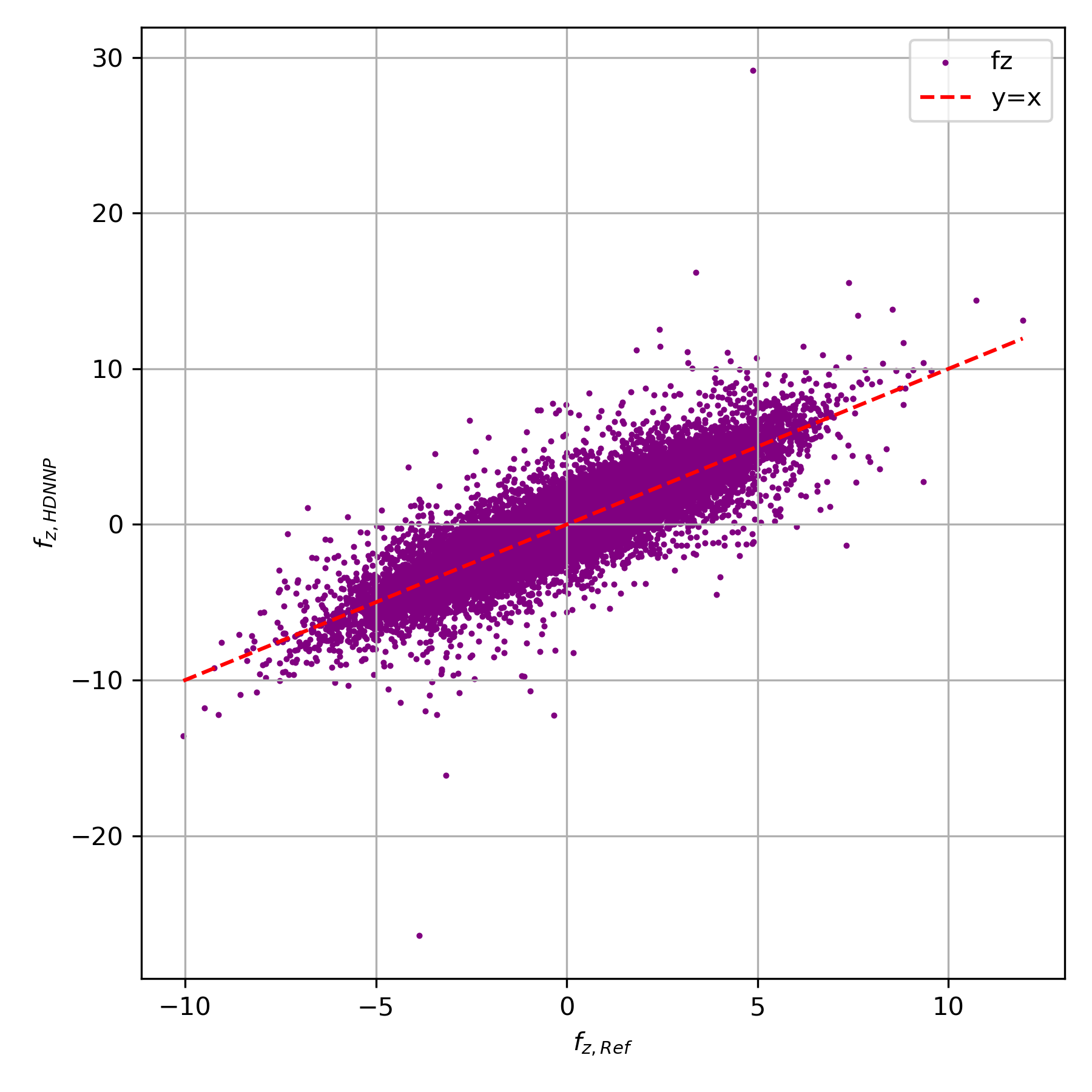} 
    \caption{AIMD (reference) z forces versus HDNNP z forces for the test set.} 
    \label{fig:Phasediagram} 
\end{figure}

\begin{figure} [h]
    \centering 
    \includegraphics[angle=0, width=0.8\textwidth]{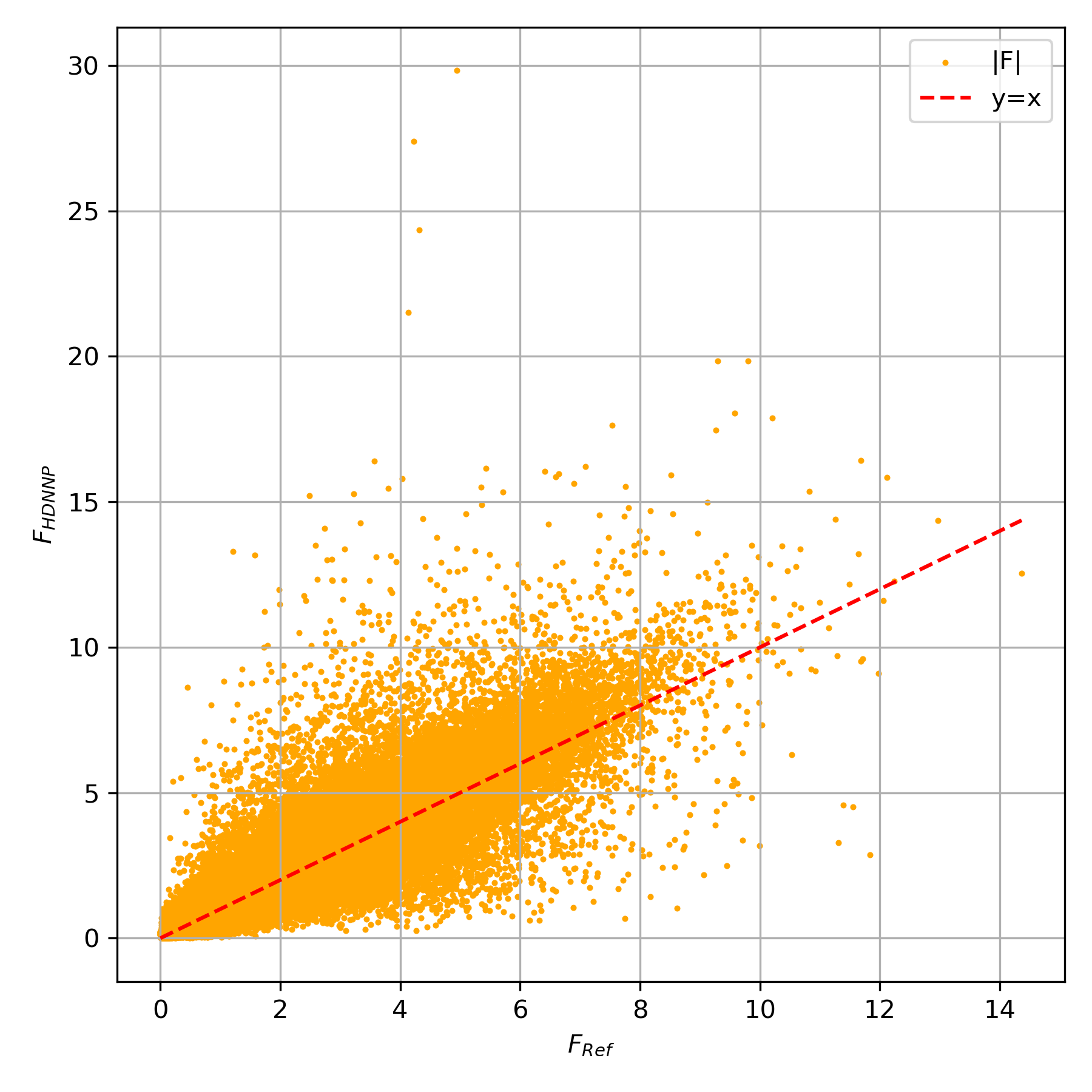} 
    \caption{AIMD (reference) forces versus HDNNP forces for the test set.} 
    \label{fig:Phasediagram} 
\end{figure}

\subsection{Structural and dynamic properties}

\begin{figure} [h]
    \centering 
    \includegraphics[angle=0, width=0.8\textwidth]{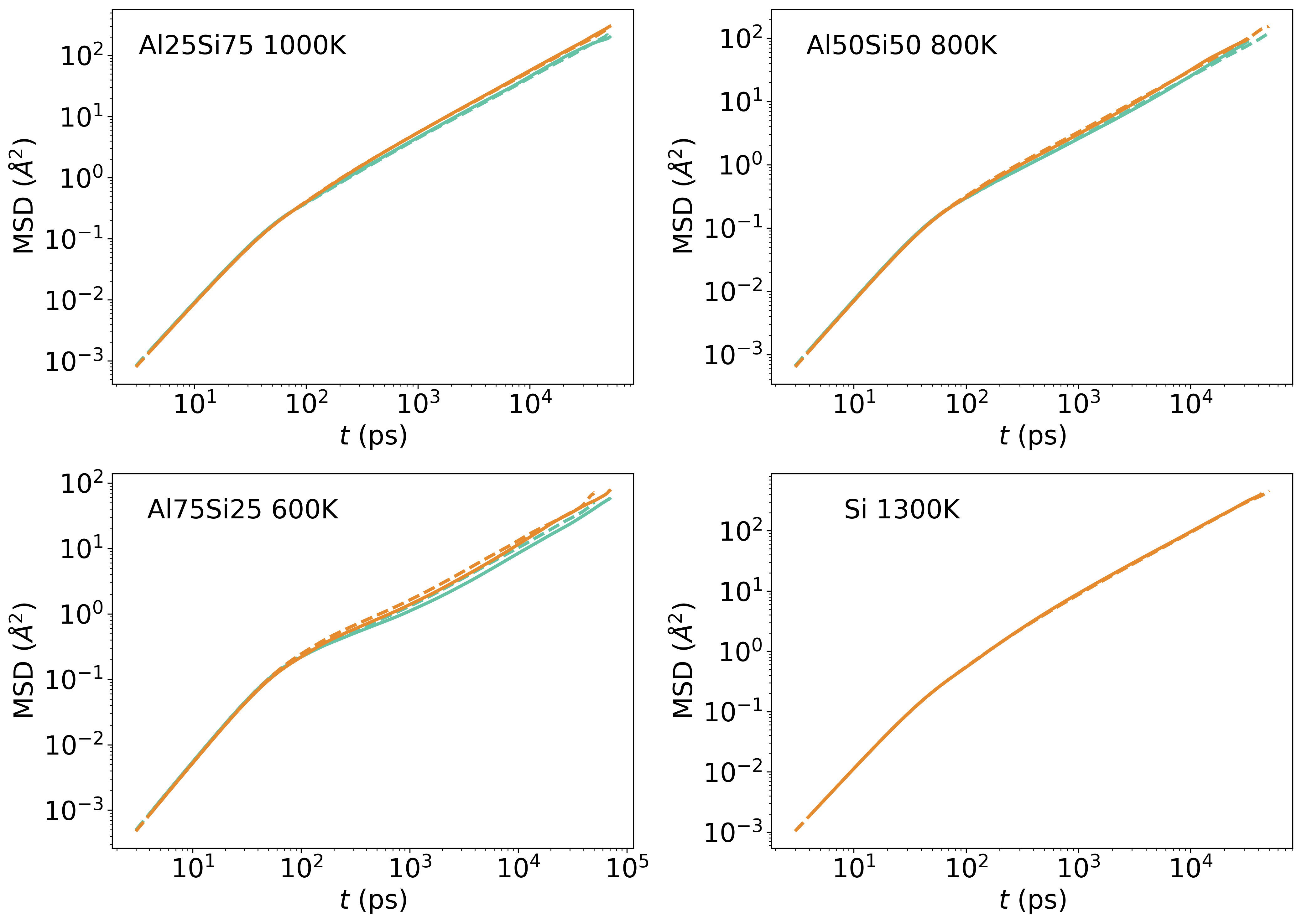} 
    \caption{Mean Square Displacements for different compositions. Dashed lines represent the HDNNP potential and solid lines represent the AIMD calculations. Orange represent Si element and green represent Al element.} 
    \label{fig:MSD} 
\end{figure}

\begin{figure} [h]
    \centering 
    \includegraphics[angle=0, width=0.8\textwidth]{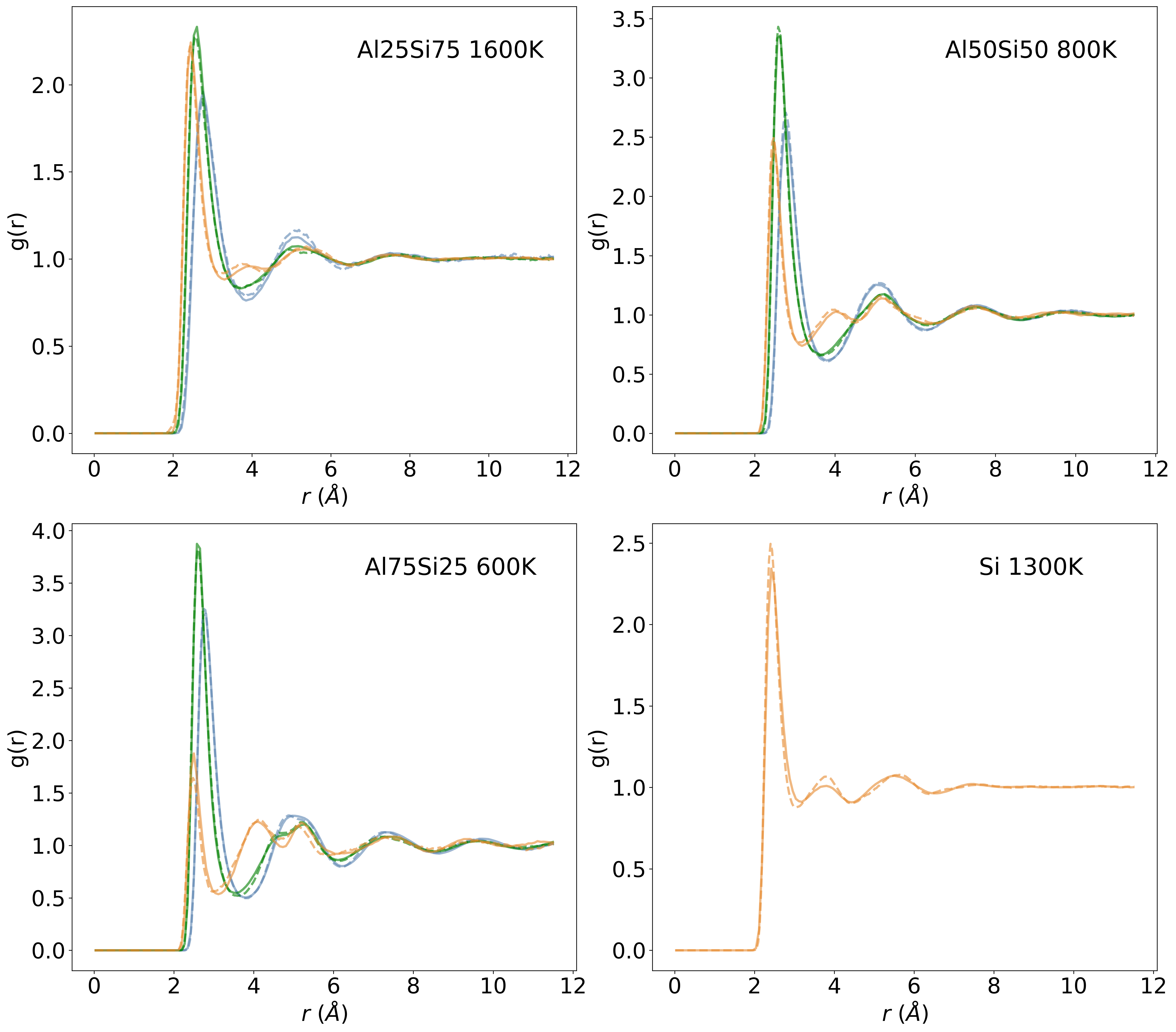} 
    \caption{Radial Distribution Function for different compositions. Dashed lines represent the HDNNP potential and solid lines represent the AIMD calculations. Orange represent Si-Si pairs, green represent Al-Al pairs, and blue represent Al-Si pairs.} 
    \label{fig:MSD} 
\end{figure}

\section{Crystal nucleation in Al-Si}
\subsection{Hyper-eutectic nucleation : case of solid Si and Al particle}

\begin{figure} [h]
    \centering 
    \includegraphics[angle=0, width=0.95\textwidth]{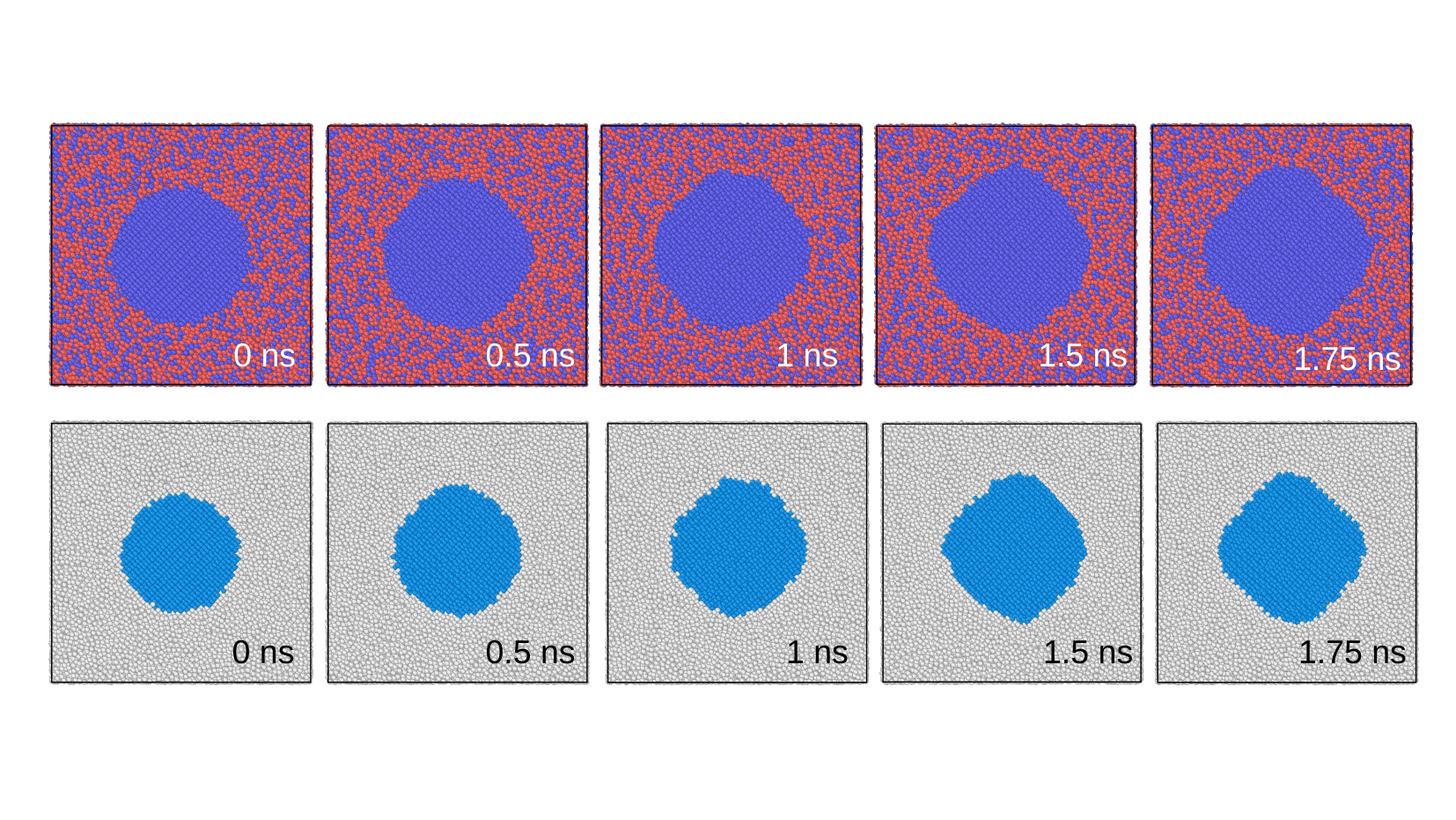} 
    \caption{Evolution of the simulation in the hyper-eutectic composition with Si crystalline particle. On top snapshots represent the evolution of the simulation with in blue Si atoms and red Al atoms. On bottom snapshots at the same time (0, 0.5, 1, 1.5, and 1.75 ns) represent the structural indicators of the atoms within the simulation. Blue represents diamond structure and white represents unknown structures (liquid or amorphous atoms). } 
    \label{fig:ParticleSigrowth} 
\end{figure}